\documentclass[conference]{IEEEtran}
\IEEEoverridecommandlockouts
\usepackage{graphicx} 
\usepackage{enumitem}
\usepackage{balance}
\usepackage{float}
\usepackage{caption}
\usepackage{hyperref}
\usepackage{amsmath}
\usepackage{amsfonts}
\usepackage{tikz}
\usetikzlibrary{bayesnet}
\usepackage{algorithm}
\usepackage{algorithmic}
\usepackage{cite}
\usepackage{amssymb}
\usepackage{textcomp}
\usepackage{xcolor}
\def\BibTeX{{\rm B\kern-.05em{\sc i\kern-.025em b}\kern-.08em
		T\kern-.1667em\lower.7ex\hbox{E}\kern-.125emX}}

\begin{document}
\title{Exploring Spatio-Temporal and Cross-Type Correlations for Crime Prediction}
\author{\IEEEauthorblockN{Xiangyu Zhao}
	\IEEEauthorblockA{Michigan State University\\
		zhaoxi35@msu.edu}
	\and
	\IEEEauthorblockN{Jiliang Tang}
	\IEEEauthorblockA{Michigan State University\\
		tangjili@msu.edu}}
\maketitle
\begin{abstract}
Crime prediction plays an impactful role in enhancing public security and sustainable development of urban. With recent advances in data collection and integration technologies, a large amount of urban data with rich crime-related information and fine-grained spatio-temporal logs has been recorded. Such helpful information can boost our understandings about the temporal evolution and spatial factors of urban crimes and can enhance accurate crime prediction. In this paper, we perform crime prediction exploiting the cross-type and spatio-temporal correlations of urban crimes. In particular, we verify the existence of correlations among different types of crime from temporal and spatial perspectives, and propose a coherent framework to mathematically model these correlations for crime prediction. The extensive experimental results on real-world data validate the effectiveness of the proposed framework. Further experiments have been conducted to understand the importance of different correlations in crime prediction.
\end{abstract}
%The vast majority of existing crime prediction algorithms either do not distinguish different types of crime or treat each crime type separately, which fail to capture the intrinsic correlations among different types of crime. 

%\vspace{-1.6mm}
\section{Introduction}
\label{sec:introduction}
It is well recognized that crime prediction is of great importance for enhancing the public security of urban so as to improve the life quality of citizens~\cite{zhao2017modeling,couch2000urban}. Accurate crime prediction is beneficial to advance the sustainable development of urban and reduce the financial loss of urban violence. Therefore, there is a rising need for precise crime prediction. Efforts have been made on constructing crime prediction models to predict either the total crime amount~\cite{zhao2017modeling,zhao2017exploring} or several specific types of crime such as \textit{Burglary}~\cite{wang2017analysis}, \textit{Felony Assault}~\cite{barrett2006predictors}, \textit{Grand Larceny}~\cite{fisher1999grand}, \textit{Murder}~\cite{revitch1978murder}, \textit{Rape}~\cite{thornhill1983human}, \textit{Robbery}~\cite{roesch1986implementation,kube1988preventing}, and \textit{Vehicle Larceny}~\cite{henry2000visualising}. In other words, most existing crime prediction methods either do not distinguish different types of crime or consider each crime type separately. 

According to criminology and recent studies, different types of crime behave differently but are intrinsically correlated. For instance, social disorganization theory~\cite{shaw1942juvenile} and broken windows theory~\cite{wilson1982broken} suggest that a series of minor crimes like vandalism or graffiti might cause the increase of more severe crimes like assaults and weapon violence; while relations between different types of crime in London are investigated in~\cite{morison2017}, and \textit{bicycle theft}, \textit{burglary}, \textit{robbery} and \textit{theft from the person} are observed to be closely related in terms of spatial distribution. The above theories and findings indicate that different types of crime are intrinsically related to each other, and exploiting the correlations among crime types could boost accurate crime prediction.

Recently, driven by the advances in big urban data collection and integration techniques, a great quantity of urban data has been collected such as crime complaint data, stop-and-frisk data, meteorological data, point of interests (POIs) data, human mobility data and 311 public-service complaint data. Such data contains rich and useful context information about crime. For example, in the near future, more crimes tend to occur in the areas with many crime complaints~\cite{zhao2017modeling}; the POIs density can characterize the neighborhood functions, which have strong impact on criminal activities according to criminal theories~\cite{brantingham1981environmental}; while public-service complaint data reveal citizens' dissatisfaction with government service, thus it is associated with crimes.  In addition, big urban data contains fine-grained information about where and when the data is collected.  Such spatio-temporal information not only enables us to study the geographical factors of crimes such as urban configuration, but also allows us to understand the dynamics and evolution of crimes over time~\cite{leong2015review}. According to environmental criminology like awareness theory~\cite{brantingham1981notes} and crime pattern theory~\cite{felson1998opportunity}, the distribution of urban crimes is highly influenced by space and time. Therefore, the spatio-temporal understandings from big urban data provide unprecedented opportunities for us to construct more accurate crime prediction. 

In this paper, we jointly explore cross-type and spatio-temporal correlations for crime prediction by leveraging big urban data. Specifically, we mainly seek answers for two challenging questions: (1) what correlations can be observed among different types of crime, and (2) how to mathematically model cross-type and spatio-temporal correlations for crime prediction. For cross-type correlations, we investigate temporal and spatial patterns of different types of crimes as well as their relationships; for spatio-temporal correlations, we focus our investigation on mathematically modeling (1) intra-region temporal correlation that suggests how crime evolves over time in a region, and (2) inter-region spatial correlation that depicts the spatial relationship across regions in the city~\cite{zhao2017modeling,zhao2017exploring,zhao2018crime}. We propose a novel framework CCC, which jointly captures \textbf{C}ross-type and spatio-temporal \textbf{C}orrelations for \textbf{C}rime prediction based on urban data.  Our major contributions can be summarized as follows: 
\vspace{-0.6mm}
\begin{itemize}[leftmargin=*]
\item We verify the existence of correlations among different types of crime from temporal and spatial perspectives; 
\item We propose a novel crime prediction framework CCC, which jointly captures cross-type and spatio-temporal correlations into a coherent model; and 
\item We conduct extensive experiments on real big urban data to validate the effectiveness of the proposed framework and the contributions of different correlations to crime prediction. 
\end{itemize}

%The rest of this paper is organized as follows. In Section 2, we formally define the problem of crime prediction. We perform preliminary data analysis in Section 3. In Section 4, we provide approaches to model cross-type and spatio-temporal correlations and introduce details about the proposed CCC framework with an optimization algorithm. Section 5 describes the dataset and presents experimental results with discussions. Section 6 briefly reviews related work. Finally, Section 7 concludes with future work.
\section{Problem Statement}
\label{sec:problem} 
In this section, we first introduce the mathematical notations and then formally define the problem we study in this work. We employ bold letters to represent vectors and matrices, e.g., $\mathbf{p}$ and $\mathbf{Q}$; we leverage non-bold letters to denote scalars, e.g., $m$ and $N$; and we use Greek letters as parameters, e.g., $\mathbf{\theta}$ and $\mathbf{\lambda}$.

Let $\mathbf{Y} \in \mathbb{R}^{N \times T \times K}$ denote the observed numbers of crime where $Y_n^t(k)$ is the number of $k^{th}$ crime type observed at $n^{th}$ region in $t^{th}$ time slot. Here we suppose that there are totally (1) $N$ regions in a city, i.e., $n = \{1, 2, \ldots, N \} \in \mathbb{R}^{N}$, (2) $T$ time slots (e.g. days, weeks, or months) in the dataset, i.e., $t = \{1, 2, \ldots, T \} \in \mathbb{R}^{T}$, and (3) $K$ types of crime (e.g. burglary, robbery and grand larceny), i.e., $k = \{1, 2, \ldots, K \} \in \mathbb{R}^{K}$. Suppose that $\mathbf{X}\in \mathbb{R}^{N \times T \times M}$ denotes the set of feature vectors, where $\mathbf{X}_n^t\in \mathbb{R}^{1 \times M}$ is the feature vector of $n^{th}$ region in $t^{th}$ time slot, and $M$ is the number of features. Note that feature vector $\mathbf{X}_n^t$ is same for all types of crime of $n^{th}$ region in $t^{th}$ time slot. More details about features will be proposed in the experiment section. 

With the above-mentioned notations and definitions, we formally state the problem of crime prediction as: \textit{Given the historical observed crime amounts $\mathbf{Y}$ and feature vectors $\mathbf{X}$, we aim to predict the crime amount of time slot $T+\tau$ (or $\tau$ time slots later) for each type of crime based on $\mathbf{Y}$ and $\mathbf{X}$.}

It should be noted that our goal is to predict crime amount for future time slot $T+\tau$. However, if the feature vector $\mathbf{X}_n^t$ is constructed based on data in $t^{th}$ time slot of $n^{th}$ region, the future feature vector of $(T+\tau)^{th}$ time slot is not available. To this end, in this paper, we actually construct $\mathbf{X}_n^t$ using data in $(t-\tau)^{th}$ time slot rather than $t^{th}$ time slot of $n^{th}$ region. Without the loss of generality, in the following sections, we leverage $\tau=1$ for illustrations, i.e., performing crime prediction for $(T+1)^{th}$ time slot. %We further assume that there is extra data where we can construct the feature matrix $\mathbf{X}^{1}$ in $t_1$. Since for different $h$ values, the major differences for the proposed framework are how to construct the feature matrices and choose the target crime amounts, in the following subsections, we will choose $h=1$ for illustrations.   
\section{Preliminary Study}
\label{sec:preliminary}
In this section, we investigate spatio-temporal and cross-type correlations for different types of crime. This preliminary analysis is based on the crime data collected from New York City, which contains 7 types of crime, i.e., Burglary, Felony Assault, Grand Larceny, Murder, Rape, Robbery, and Vehicle Larceny. We will study cross-type correlations from temporal and spatial perspectives.

\begin{figure}[H]
	\centering
	\includegraphics[width=90mm]{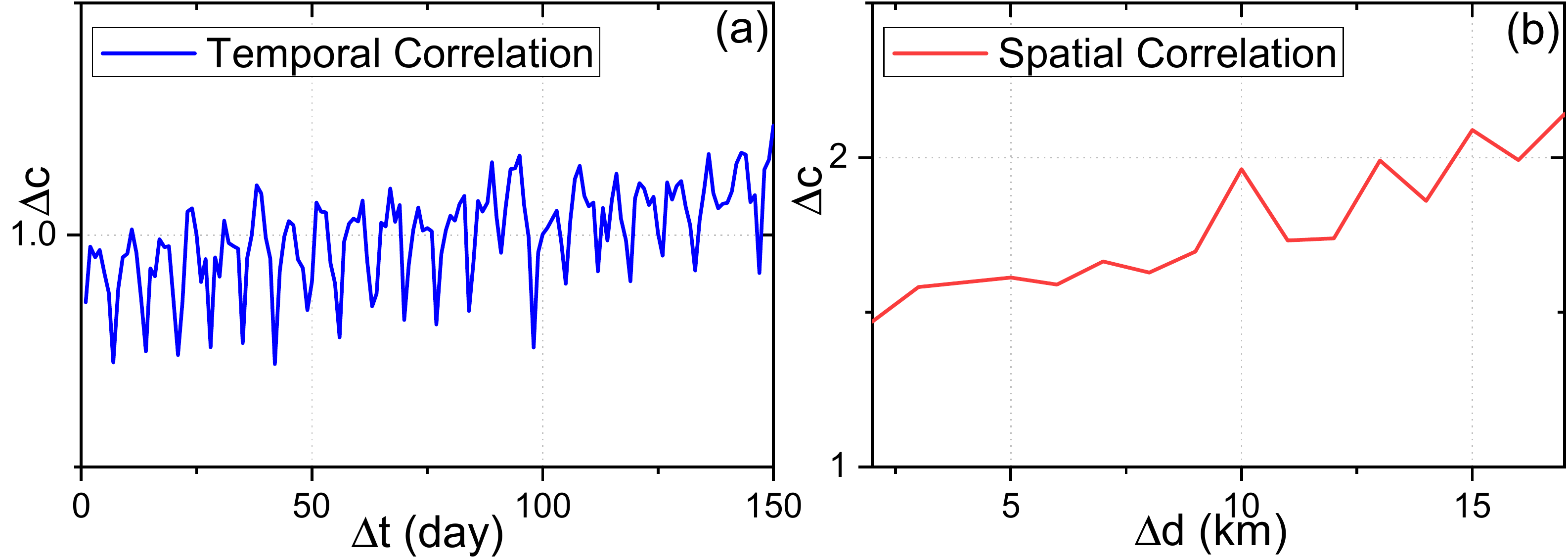}
	\caption{The temporal and spatial correlations of grand larceny.}
	\label{fig:Fig_Preliminary_ST}
\end{figure} 
\subsection{Spatio-Temporal Correlations}
\label{sec:spatio-temporal}
Within a region, the amount of crime should change smoothly over time. We assume the crime amount is $c_t$ and $c_{t+\Delta t}$ for time $t$ and $t+ \Delta t$. To study temporal correlation, we show how the crime amount differences $|c_t - c_{t+\Delta t} |$ changes with $\Delta t$ on average of all regions. The result is illustrated in Figure~\ref{fig:Fig_Preliminary_ST}(a) where x-axis is $\Delta t$ (days) and y-axis is $|c_t - c_{t+\Delta t} |$. From Figure~\ref{fig:Fig_Preliminary_ST}(a), we can observe that the crime differences are highly related to $\Delta t$. To be specific, (i) two consecutive time slots share similar crime amounts; (ii) with the increase of $\Delta t$, the crime difference is likely to increase.

For regions in the city, if two regions are spatially close to each other, they are likely to have similar crime amounts at the same time slot. Given a pair of regions, we leverage $\Delta d$ as their spatial distance and use $\Delta c$ as their absolute crime difference. We show how $\Delta c$ changes with $\Delta d$ averaged over all time slots in Figure~\ref{fig:Fig_Preliminary_ST}(b), where x-axis is $\Delta d$ and y-axis is $\Delta c$. We note that (i) when two regions are spatially close, they have similar crime amounts and (ii) with the increase of distance $\Delta d$, the crime difference $\Delta c$ tends to increase. 

The above observations suggest the existence of temporal and spatial correlations for each type of urban crime. Note that we illustrate the observation of grand larceny in Figure~\ref{fig:Fig_Preliminary_ST}, while omit other types of crime which have similar observations.

\begin{figure}[H]
	\centering
	\includegraphics[width=90mm]{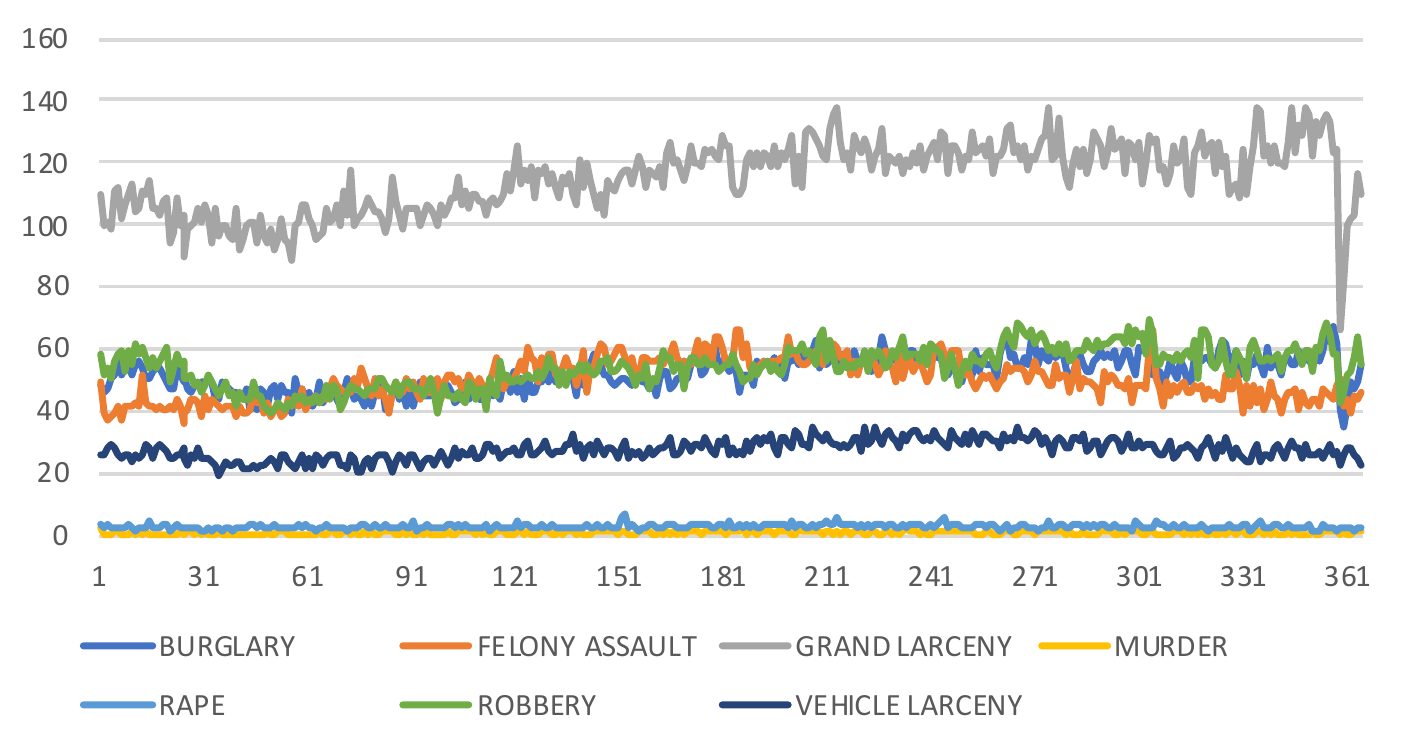}
	\caption{Average daily crime numbers from 2006 to 2015.}
	\label{fig:Preliminary_T}
\end{figure}

\subsection{Cross-Type Correlations}
\label{sec:cross-type}
To investigate temporal correlations among different types of crime, we study how crime amounts of each type change with the days of a year. The average daily crime amounts from 2006 to 2015 are shown in Figure~\ref{fig:Preliminary_T}, where x-axis denotes the days of a year and y-axis is the crime amounts for each type of crime, respectively. From the figure, we observe temporal correlations between different types of crime. Specifically, the daily crime amounts of most types tend to increase from March to September and decrease from October to February. Furthermore, the crime amounts of some types such as Burglary, Grand Larceny and Robbery tend to increase before Christmas, but decrease dramatically during Christmas and New Year.

\begin{figure}[t]
	\centering
	\includegraphics[width=85mm]{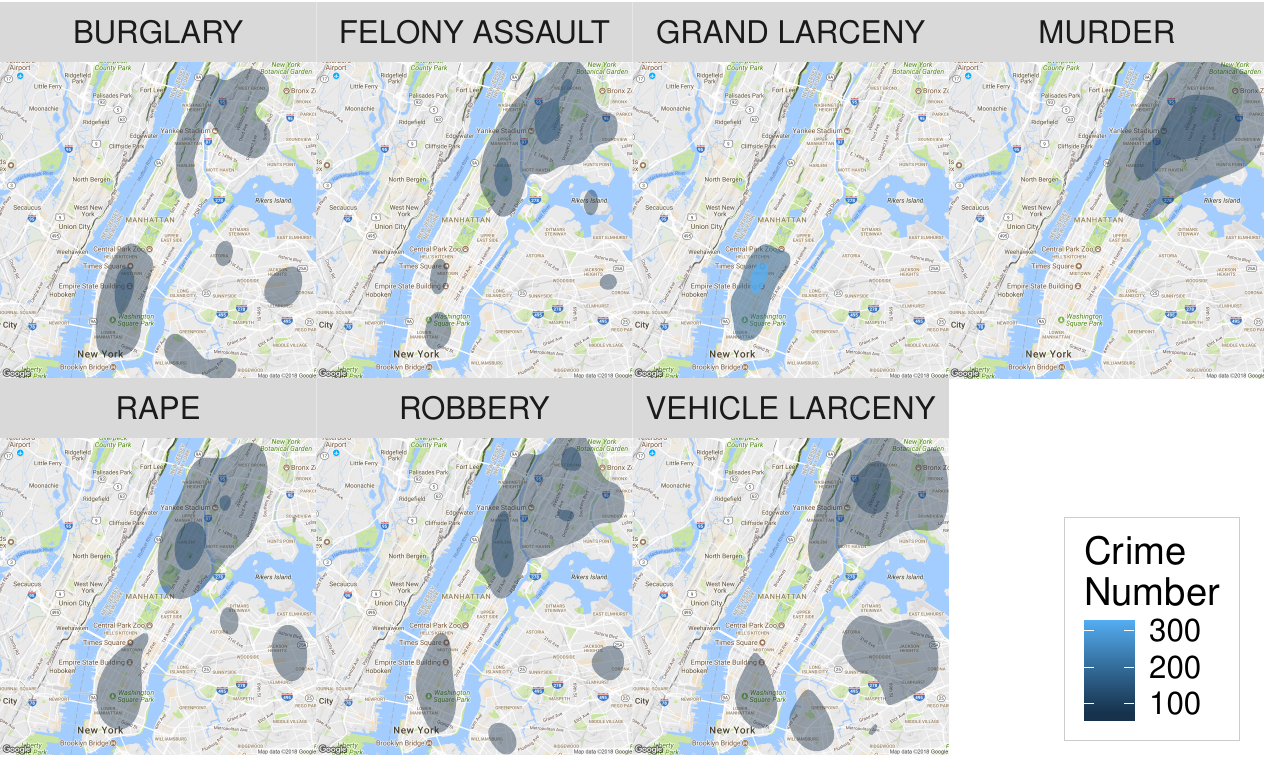}
	\caption{Spatial distribution of crimes in 2012.}
	\label{fig:Preliminary_S}
%	\vspace{-3.6mm}
\end{figure}

To study the spatial correlations among different types of crime, we show how crimes spatially distribute in New York City of 2012 in Figure~\ref{fig:Preliminary_S}. From Figure~\ref{fig:Preliminary_S}, we make the observations that (1) the majority of types concentrate in the Bronx, except for Grand Larceny; (2) Manhattan is also a hot district for some types, especially for Grand Larceny and Burglary; and (3) Burglary, Rape, Robbery, and Vehicle Larceny share some hotspots in the Brooklyn and Queens.

\begin{figure}[t]
	\centering
	\includegraphics[width=72mm]{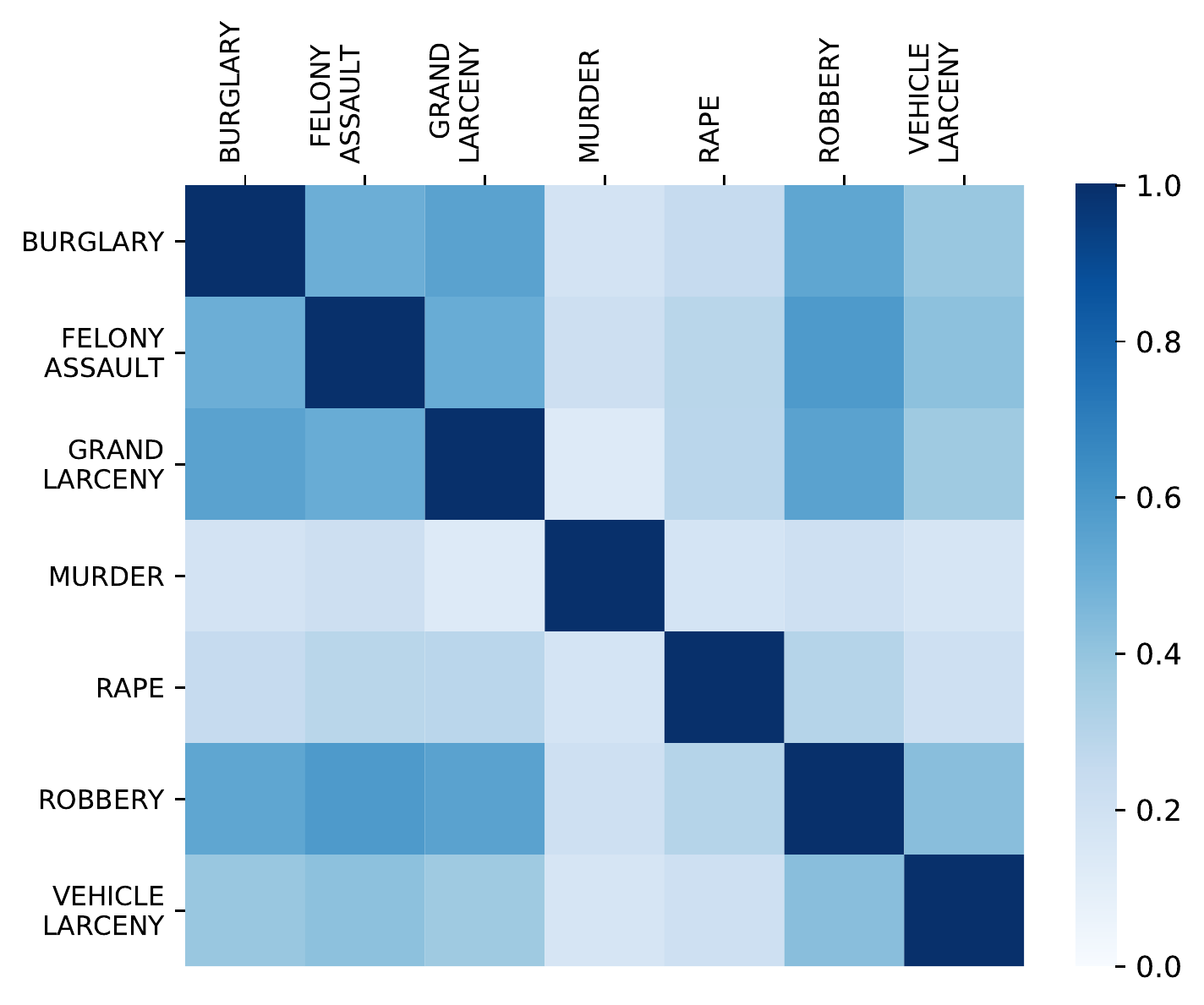}
	\caption{Cross-type temporal-spatial correlation matrix heatmap.}
	\label{fig:Preliminary_ST}
	%	\vspace{-3.6mm}
\end{figure}

To study the correlations from both temporal and spatial perspective, we first construct $K = 7$ matrices $\{\mathbf{R}^1, \mathbf{R}^2, \ldots, \mathbf{R}^K\}$, where each $\mathbf{R}^k \in \mathbb{R}^{N \times T}$. Each element $R_{n,t}^k \in \mathbf{R}^k$ is the crime amount for $k^{th}$ type of crime in $n^{th}$ region of $t^{th}$ time slot. To study the spatio-temporal correlations between two types (e.g. the $i^{th}$ and $j^{th}$ type) of crime, we calculate the variant of \textit{cosine similarity} between $\mathbf{R}^i$ and $\mathbf{R}^j$ as follows:
\begin{small}
\begin{align}
cosine(\mathbf{R}^i,\mathbf{R}^j)=\frac{<\mathbf{R}^i,\mathbf{R}^j>}{\|\mathbf{R}^i\|_F \|\mathbf{R}^j\|_F},
\end{align}
\end{small}
where $<\mathbf{R}^i,\mathbf{R}^j> = \sum_{n,t}R_{n,t}^i R_{n,t}^j$ and $\|\cdot\|_F$ is the \textit{Frobenius norm}. The result is shown in Figure \ref{fig:Preliminary_ST}. We can observe that most types of crime are indeed correlated with each other. The least spatio-temporal correlation exists between Grand Larceny and Murder, which is also demonstrated in Figure~\ref{fig:Preliminary_T} and Figure~\ref{fig:Preliminary_S}.

To sum up, we demonstrate the existence of temporal and spatial correlations among different types of crime. These observations provide the groundwork for us to leverage the cross-type correlations for accurate crime prediction.

\section{The Proposed Crime Prediction Framework}
\label{sec:framework}
In above section, we validate the correlations among different types of crime. In this section, we will first present the basic model without considering cross-type and spatio-temporal correlations, then propose the details of introducing cross-type correlations as well as spatio-temporal correlations into a coherent framework. Finally, we will discuss the optimization process of the proposed framework and how to leverage the framework to perform crime prediction.  

\subsection{The Basic Model}
Without considering cross-type and spatio-temporal correlations, we build a basic and individual model of $k^{th}$ crime type for $n^{th}$ region in $t^{th}$ time slot. Correspondingly, there is a weight vector $\mathbf{W}_n^t(k) \in \mathbb{R}^{M \times 1}$ for $k^{th}$ crime type of $n^{th}$ region in $t^{th}$ time slot, which can map $\mathbf{X}_n^t$ to  $Y_n^t(k)$ as: $\mathbf{X}_n^t \mathbf{W}_n^t(k) \rightarrow Y_n^t(k)$. All $\mathbf{W}_n^t(k)$ can be learned by solving the following regression problem: 
\begin{small}
\begin{align}\label{equ:loss_function}
	\min_{\mathbf{W}_n^t(k)} \sum_{n=1}^N \sum_{t=1}^T \sum_{k=1}^K \left( \left(\mathbf{X}_n^t \mathbf{W}_n^t(k) - Y_n^t(k)\right)^2 + \theta \| \mathbf{W}_n^t(k) \|_2^2 \right),
\end{align}
\end{small}
where the first term is the square loss function for regression task in this work. Note that it is straightforward to leverage other loss functions such as logistic loss and hinge loss. We employ $\| \mathbf{W}_n^t(k) \|_2^2$ (controlled by a non-negative parameter $\theta$) to avoid over-fitting issue. This basic and individual model completely neglects the existence of correlations among different types of crime and spatio-temporal correlations within each type of crime. In the following subsections, we will discuss how to model cross-type correlations as well as spatio-temporal correlations based on this basic model.  

\subsection{Modeling Cross-Type Correlations}
\label{sec:type_correlation}
Our preliminary study Section~\ref{sec:cross-type} validates the existence of correlations among different types of crime. In this subsection, we will introduce the model component to capture cross-type correlations.

To exploit correlations of urban crimes, we first decompose the weight vector $\mathbf{W}_n^t(k)$ into the sum of two components $\mathbf{W}_n^t(k) = \mathbf{P}_n^t + \mathbf{Q}_n^t(k)$, where we use $\mathbf{P}_n^t$ to capture the common features shared by all crime types of $n^{th}$ region in $t^{th}$ time slot, while $\mathbf{Q}_n^t(k)$ captures the specific features for $k^{th}$ crime type. For instance, some common features lead to the concentration of most crime types in Bronx, while some specific features cause the Grand Larceny concentrating in Manhattan. We will leverage different regularization terms on $\mathbf{P}$ and $\mathbf{Q}$ to exploit different correlations.

$\mathbf{Q}_n^t(k)$ can represent the $k^{th}$ crime type, which paves us a way to capture cross-type correlations. We first combine all the type specific weight vectors into a weight matrix, i.e., $\mathbf{Q}_n^t = [\mathbf{Q}_n^t(1), \mathbf{Q}_n^t(2), \ldots,$ $ \mathbf{Q}_n^t(K)]\in \mathbb{R}^{M \times K}$. Then, adopting the task relationship regularization component in~\cite{zhang2012convex}, the relationships among $\mathbf{Q}_n^t(1), \mathbf{Q}_n^t(2), \ldots,$ $ \mathbf{Q}_n^t(K)$ can be modeled as as follows:
\begin{small}
\begin{equation}\label{eq:type}
\begin{aligned}
&\sum_{n=1}^N \sum_{t=1}^T \alpha \cdot tr\left(\mathbf{Q}_n^t {\mathbf{\Omega}_n^t}^{-1} \mathbf{Q}_n^{t\top}\right),\\
&\,\,\,\,\,\,\,\,\,\,\,s.t. \,\,\,\,\,\,\,\,\,\,\mathbf{\Omega}_n^t \geq 0\\
&\,\,\,\,\,\,\,\,\,\,\,\,\,\,\,\,\,\,\,\,\,\,\,\,\,\,\,\,\, tr\left(\mathbf{\Omega}_n^t\right) = K
\end{aligned}
\end{equation}
\end{small}
where $\mathbf{\Omega}_n^t$ is the crime type covariance matrix of $n^{th}$ region in $t^{th}$ time slot to learn and $\alpha$ is a non-negative parameter to control the contributions by exploring cross-type correlations. Since $\mathbf{\Omega}_n^t$ is a covariance matrix, the matrix $\mathbf{\Omega}_n^t$ should be positive semidefinite (or $\mathbf{\Omega}_n^t \geq 0$). We introduce this regularization component to capture the correlations among different type of crimes of $n^{th}$ region in $t^{th}$ time slot based on $\mathbf{Q}_n^t$ and $\mathbf{\Omega}_n^t$.

\subsection{Modeling Intra-Region Temporal Correlation}
\label{sec:temporal_correlation}
Crime within a region is observed following intra-region temporal correlation in Section~\ref{sec:spatio-temporal} -- (1) for two consecutive time slots, they tend to share similar crime amounts; and (2) with the increase of distance between two time slots, the crime amounts difference is likely to increase. Inspired by this discovery, we propose a temporal regularization component to model the temporal correlations of crime amount within each region. 

To be specific, considering the smooth evolution of crime amounts, the weight vectors should also change smoothly. Therefore, we adopt a series of discrete weight vectors over time to represent the temporal dynamics of crime amounts, and we add a temporal regularization component to the basic model as follows:
\begin{small}
\begin{equation}\label{eq:temporal}
\beta \cdot \sum_{n=1}^N {\sum_{t=1}^{T-1} \bigg(\|\mathbf{P}_n^t - \mathbf{P}_n^{t+1}\|_1 + \sum_{k=1}^{K}\|\mathbf{Q}_n^t(k) - \mathbf{Q}_n^{t+1}(k)\|_1\bigg)},    
\end{equation}
\end{small}
where non-negative parameter $\beta$ is introduced to control the contribution of intra-region temporal correlation from the temporal regularization component. The first term pushes $\mathbf{P}_n^t$ as closer as $\mathbf{P}_n^{t+1}$, i.e., the weight vector for common features shared by all crime types of $n^{th}$ region change smoothly over time, while the second term captures the smooth evolution of weight vector for each specific crime type within a region. Note that we define $\| {\bf X} \|_1$ as $\sum_{i,j} |{\bf X}_{ij}|$ in this work, which makes it possible to encourage weight vectors of two consecutive time slots to be exactly same. We do not use $\ell_2$-norm since it is likely cause ``wiggly'' cost dynamics, which is not robust to noises and may hurt generalization performance~\cite{zheng2013time}. Eq.~(\ref{eq:temporal}) can be rewritten as:
%\begin{equation}\label{eq:fuse_lasso}
%\begin{aligned}
%\beta \cdot \sum_{n=1}^N \sum_{t=1}^{T-1} & \bigg(\|\mathbf{P}_n^t - \mathbf{P}_n^{t+1}\|_1 + \sum_{k=1}^{K}\|\mathbf{Q}_n^t(k) - \mathbf{Q}_n^{t+1}(k)\|_1\bigg)\\
%= \sum_{n=1}^N&\bigg( \|\mathbf{P}_{n}\mathbf{A}\|_1 + \sum_{k=1}^{K}\|\mathbf{Q}_n(k)\mathbf{A}\|_1\bigg),
%\end{aligned}
%\end{equation}
\begin{small}
\begin{equation}\label{eq:fuse_lasso}
\begin{aligned}
\sum_{n=1}^N\bigg( \|\mathbf{P}_{n}\mathbf{A}\|_1 + \sum_{k=1}^{K}\|\mathbf{Q}_n(k)\mathbf{A}\|_1\bigg),
\end{aligned}
\end{equation}
\end{small}
where $\mathbf{P}_{n} = [\mathbf{P}_n^1, \mathbf{P}_n^2, \ldots, \mathbf{P}_n^T]\in \mathbb{R}^{M \times T}$ and $\mathbf{Q}_{n}(k)= [\mathbf{Q}_n^1(k), \mathbf{Q}_n^2(k),$ $\ldots, \mathbf{Q}_n^T(k)]\in \mathbb{R}^{M \times T}$. $\mathbf{A}\in \mathbb{R}^{T\times(T-1)}$ is a sparse matrix. More specifically, $\mathbf{A}(t, t)=\beta, \mathbf{A}(t+1, t)= -\beta$ for $t=1, \ldots, T-1$ and all the other terms 0. 

\begin{figure}[t]
	\centering
	\includegraphics[width=100mm]{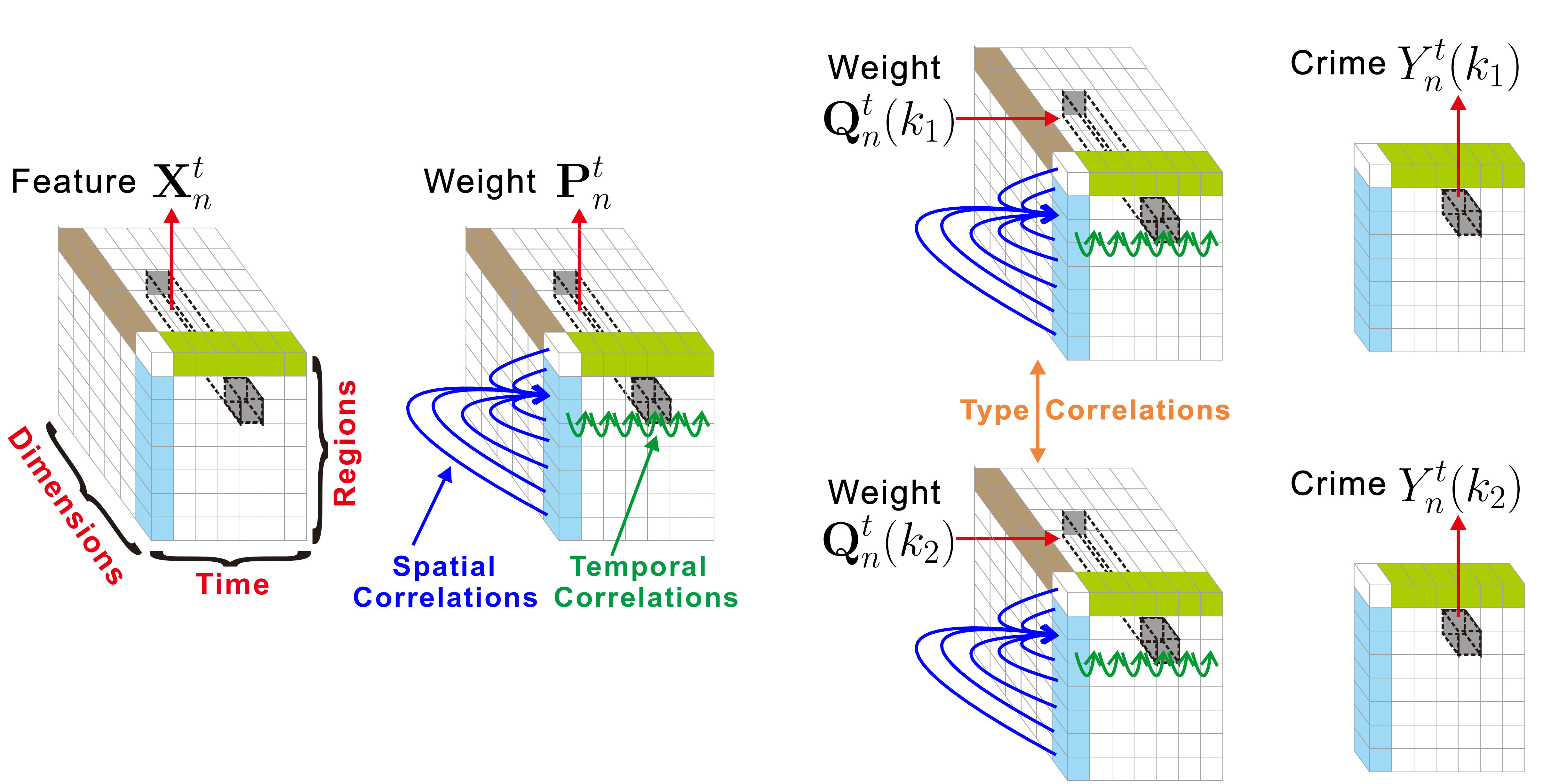}
	\caption{An illustration of the proposed framework with two types of crime.\label{fig:st}}
%	\vspace{-3.6mm}
\end{figure}

\subsection{Modeling Inter-Region Spatial Correlation}
\label{sec:spatial_correlation}
As mentioned in Section~\ref{sec:spatio-temporal}, aside from intra-region temporal correlation, the crime amounts across all regions follow inter-region spatial correlation -- (1) two spatial close regions tend to have similar crime amounts; and (2) with the increase of geographical distance between two regions in a city, the crime difference between these two regions is likely to increase in a certain time slot. This observation inspires us to develop a spatial regularization component to capture the spatial correlation of crime amounts across regions in a city. 

Specifically, we choose to minimize the following spatial component to capture inter-region spatial correlation:
\begin{small}
\begin{equation}
	\sum_{t=1}^T \sum_{i=1}^N \sum_{j=1}^N d(i,j)^{-\gamma}\bigg(\|\mathbf{P}_i^t- \mathbf{P}_j^t\|_1 + \sum_{k=1}^K \|\mathbf{Q}_i^t(k)- \mathbf{Q}_j^t(k)\|_1 \bigg),
    \label{equ:spatial_smoothess}
\end{equation}
\end{small}
where $d(i,j)$ is the spatial distance between $i^{th}$ and $j^{th}$ region. $d(i,j)^{-\gamma}$ is a power law exponential function, which is non-increase in terms of $d(i,j)$, where $\gamma$ is the parameter controlling the degree of spatial correlations. Thus, when $i^{th}$ and $j^{th}$ regions are closer, (i.e. $d(i,j)$ is smaller), $d(i,j)^{-\gamma}$ becomes larger that enforces weight vectors of two regions to be closer. Similar analysis can be used when the distance between $i^{th}$ and $j^{th}$ is larger. 

Similar to intra-region temporal correlation, the first term pushes $\mathbf{P}_i^t$ and $\mathbf{P}_j^t$ to be closer, which means the weight vector for common features of all types of crime in $i^{th}$ and $j^{th}$ region is similar if they are spatially close to each other. The second term captures the proximity across regions of each type of crime. This spatial component encodes Tobler's first law of geography~\cite{tobler1970computer} and performs a soft constraint that spatially close regions tend to have similar weight vectors. We can rewrite Eq.~(\ref{equ:spatial_smoothess}) as:
%\begin{equation}
%\begin{aligned}
%	\sum_{t=1}^T \sum_{i=1}^N \sum_{j=1}^N& d(i,j)^{-\gamma}\bigg(\|\mathbf{P}_i^t- \mathbf{P}_j^t\|_1 + \sum_{k=1}^K \|\mathbf{Q}_i^t(k)- \mathbf{Q}_j^t(k)\|_1 \bigg)\\
%	= \sum_{t=1}^T&\bigg( \| \mathbf{P}^{t}\mathbf{B}\|_1 + \sum_{k=1}^K \| \mathbf{Q}^{t}(k)\mathbf{B}\|_1 \bigg),
%	\label{equ:graph_lasso}\\
%\end{aligned}
%\end{equation}
\begin{small}
\begin{equation}
\begin{aligned}
\sum_{t=1}^T\bigg( \| \mathbf{P}^{t}\mathbf{B}\|_1 + \sum_{k=1}^K \| \mathbf{Q}^{t}(k)\mathbf{B}\|_1 \bigg),
\label{equ:graph_lasso}\\
\end{aligned}
\end{equation}
\end{small}
where $\mathbf{P}^t = [\mathbf{P}_1^t, \mathbf{P}_2^t, \ldots, \mathbf{P}_N^t]\in \mathbb{R}^{M \times N}$ and $\mathbf{Q}^t(k) = [\mathbf{Q}_1^t(k), \mathbf{Q}_2^t(k),$ $\ldots, \mathbf{Q}_N^t(k)]\in \mathbb{R}^{M \times N}$. $\mathbf{B}\in \mathbb{R}^{N\times N^{2}}$ is a sparse matrix. To be specific, we have $\mathbf{P}(i, (i-1) \cdot  N+j)=d(i,j)^{-\gamma}$ and $\mathbf{P}(j, (i-1) \cdot  N+j)=-d(i,j)^{-\gamma}$ for $i=1, \ldots, N, j=1, \ldots , N$ and $i\not= j$, while all the other terms 0.

\subsection{An Optimization Method}
\label{sec:optimization}
With aforementioned components to capture cross-type correlations and spatio-temporal correlations, the objective loss function of the proposed framework is to solve the following optimization task: 
\begin{small}
\begin{equation}\label{equ:loss_function1}
\begin{aligned}
	\min_{\mathbf{P,Q,\Omega}}L & = \sum_{n=1}^N \sum_{t=1}^T \sum_{k=1}^K \left(\mathbf{X}_n^t \left(\mathbf{P}_n^t + \mathbf{Q}_n^t(k)\right) - Y_n^t(k)\right)^2\\
	& + \sum_{n=1}^N \sum_{t=1}^T \alpha \cdot tr\left(\mathbf{Q}_n^t {\mathbf{\Omega}_n^t}^{-1} \mathbf{Q}_n^{t\top}\right)\\
	& + \sum_{n=1}^N\bigg( \|\mathbf{P}_{n}\mathbf{A}\|_1 + \sum_{k=1}^{K}\|\mathbf{Q}_n(k)\mathbf{A}\|_1\bigg)\\
	& + \sum_{t=1}^T\bigg( \| \mathbf{P}^{t}\mathbf{B}\|_1 + \sum_{k=1}^K \| \mathbf{Q}^{t}(k)\mathbf{B}\|_1 \bigg),\\
	s.t. \,\,\,\,&\,\,\,\,\,\,\,\,\mathbf{\Omega}_n^t \geq 0 \,\,\,\,\,\,\,\,\,\,\,\,\,\,\,\,\,\,\,\,\,\,\,\forall n \in [1,N]\,\,\,\,\,\,\forall t \in [1,T]\\
	\,\,\,\,\,\,\,\,\,&\,\,\,\,\,\,\,\, tr\left(\mathbf{\Omega}_n^t\right) = K\,\,\,\,\,\,\,\,\,\forall n \in [1,N]\,\,\,\,\,\,\forall t \in [1,T]
\end{aligned}
\end{equation} 
\end{small}
where first term is the basic regression model, the second term captures cross-type correlations, the third term models intra-region temporal correlations and the last term captures the inter-region spatial correlations.  Figure~\ref{fig:st} is an illustration of the proposed framework with two types of crime, where orange arrows are for cross-type correlations between two types of crime, green arrows are for temporal correlations and blue arrows are for spatial correlations. 

In this work, we leverage ADMM technique~\cite{boyd2011distributed} to optimize the objective loss function Eq.~(\ref{equ:loss_function1}). We first suppose $\mathbf{C}_{n}=\mathbf{P}_{n}\mathbf{A}\in \mathbb{R}^{M \times T-1}$, $ \mathbf{D}_{n}(k)=\mathbf{Q}_{n}(k)\mathbf{A}\in \mathbb{R}^{M \times T-1}$, $\mathbf{E}^{t}=\mathbf{P}^{t}\mathbf{B}\in \mathbb{R}^{M \times N^2}$ and $\mathbf{F}^{t}(k)=\mathbf{Q}^{t}(k)\mathbf{B}\in \mathbb{R}^{M \times N^2}$, where $\mathbf{C}_{n}$, $\mathbf{D}_{n}(k)$, $\mathbf{E}^{t}$ and $\mathbf{F}^{t}(k)$ are \textit{auxiliary variable matrices} in ADMM. Then the objective loss function becomes:
\begin{small}
\begin{equation}\label{equ:loss_function2}
\begin{aligned}
\min_{\mathbf{P,Q,\Omega}}L & = \sum_{n=1}^N \sum_{t=1}^T \sum_{k=1}^K \left(\mathbf{X}_n^t \left(\mathbf{P}_n^t + \mathbf{Q}_n^t(k)\right) - Y_n^t(k)\right)^2\\
& + \sum_{n=1}^N \sum_{t=1}^T \alpha \cdot tr\left(\mathbf{Q}_n^t {\mathbf{\Omega}_n^t}^{-1} \mathbf{Q}_n^{t\top}\right)\\
& + \sum_{n=1}^N \bigg(\|\mathbf{C}_{n}\|_1 + \sum_{k=1}^{K}\|\mathbf{D}_n(k)\|_1\bigg)\\
& + \sum_{t=1}^T\bigg( \| \mathbf{E}^{t}\|_1 + \sum_{k=1}^K \| \mathbf{F}^{t}(k)\|_1\bigg),\\
s.t. \,\,\,\,&\,\,\,\,\,\,\,\,\mathbf{\Omega}_n^t \geq 0\,\,\,\,\,\,\,\,\,\,\,\,\,\,\,\,\,\,\,\,\,\,\,\,\,\,tr\left(\mathbf{\Omega}_n^t\right) = K\\
\,\,\,\,\,\,\,\,\,&\,\,\,\,\,\,\,\,\mathbf{C}_{n}=\mathbf{P}_{n}\mathbf{A}\,\,\,\,\,\,\,\,\,\,\,\,\,\,\,\,\mathbf{D}_{n}(k)=\mathbf{Q}_{n}(k)\mathbf{A}\\
\,\,\,\,\,\,\,\,\,&\,\,\,\,\,\,\,\,\mathbf{E}^{t}=\mathbf{P}^{t}\mathbf{B}\,\,\,\,\,\,\,\,\,\,\,\,\,\,\,\,\,\,\,\mathbf{F}^{t}(k)=\mathbf{Q}^{t}(k)\mathbf{B}\\
\,\,\,\,\,\,\,\,\,&\,\,\,\,\,\,\,\,\forall n \in [1,N]\,\,\,\,\,\forall t \in [1,T]\,\,\,\,\,\forall k \in [1,K]
\end{aligned}
\end{equation} 
\end{small}
Then the scaled form of ADMM optimization formulation of Eq~(\ref{equ:loss_function2}) can be written as:
\begin{small}
\begin{equation}\label{equ:loss_admm}
\begin{aligned}
	&\min L_{\rho}(\mathbf{P},\mathbf{Q},\mathbf{\Omega},\mathbf{C},\mathbf{D},\mathbf{E},\mathbf{F},\mathbf{S},\mathbf{U},\mathbf{V},\mathbf{Z})\\
	&=\sum_{n=1}^N \sum_{t=1}^T \sum_{k=1}^K \left(\mathbf{X}_n^t \left(\mathbf{P}_n^t + \mathbf{Q}_n^t(k)\right) - Y_n^t(k)\right)^2 \\
	&+\sum_{n=1}^N \sum_{t=1}^T \alpha \cdot tr\left(\mathbf{Q}_n^t {\mathbf{\Omega}_n^t}^{-1} \mathbf{Q}_n^{t\top}\right)\\
	&+\sum_{n=1}^N \left(\| \mathbf{C}_{n}\|_1 + \frac{\rho}{2} \| \mathbf{P}_{n}\mathbf{A} -\mathbf{C}_{n} + \mathbf{S}_{n}\|_F^2\right)\\
	&+\sum_{n=1}^N\sum_{k=1}^K \left(\| \mathbf{D}_{n}(k)\|_1 + \frac{\rho}{2} \| \mathbf{Q}_{n}(k)\mathbf{A} -\mathbf{D}_{n}(k) + \mathbf{U}_{n}(k)\|_F^2\right)\\
	&+\sum_{t=1}^T\left(\| \mathbf{E}^{t}\|_1 + \frac{\rho}{2}\| \mathbf{P}^{t}\mathbf{B} - \mathbf{E}^{t} + \mathbf{V}^{t} \|_F^2\right)\\
	&+\sum_{t=1}^T\sum_{k=1}^K\left(\| \mathbf{F}^{t}(k)\|_1 + \frac{\rho}{2}\| \mathbf{Q}^{t}(k)\mathbf{B} - \mathbf{F}^{t}(k) + \mathbf{Z}^{t}(k) \|_F^2\right)\\
	& s.t. \,\,\,\,\,\,\,\,\,\,\,\,\mathbf{\Omega}_n^t \geq 0\,\,\,\,\,\,\,\,\,\,\,\,\,\,\,\,\,\,\,\,\,\,\,\,\,\,\,tr\left(\mathbf{\Omega}_n^t\right) = K\\
	&\,\,\,\,\,\,\,\,\,\,\,\,\,\,\,\,\,\,\,\,\,\forall n \in [1,N]\,\,\,\,\,\,\,\,\,\,\,\,\,\,\,\,\,\forall t \in [1,T]
\end{aligned}
\end{equation}
\end{small}
where $\|\cdot\|_F$ is the \textit{Frobenius}-norm of a matrix. We introduce \textit{scaled dual variable matrices} $\mathbf{S}_{n} \in \mathbb{R}^{M\times (T-1)}$, $\mathbf{U}_{n}(k) \in \mathbb{R}^{M\times (T-1)}$, $\mathbf{V}^{t} \in \mathbb{R}^{M\times N^{2}}$ and $\mathbf{Z}^{t}(k) \in \mathbb{R}^{M\times N^{2}}$ of ADMM. The penalty for the violation of equality constraints $\mathbf{C}_{n}=\mathbf{P}_{n}\mathbf{A}$, $\mathbf{D}_{n}(k)=\mathbf{Q}_{n}(k)\mathbf{A}$, $\mathbf{E}^{t}=\mathbf{P}^{t}\mathbf{B}$, $\mathbf{F}^{t}(k)=\mathbf{Q}^{t}(k)\mathbf{B}$ is controlled by a non-negative parameter $\rho$. According to ADMM technique, each optimization iteration of Eq~(\ref{equ:loss_admm}) consists of the following steps:
\begin{small}
\begin{align}  
	&\mathbf{P}_n^t \leftarrow \mathbf{P}_n^t - \eta\frac{\partial L_{\rho}}{\partial \mathbf{P}_n^t},\label{equ:update_P}\\ 
	&\mathbf{Q}_n^t(k) \leftarrow \mathbf{Q}_n^t(k) - \eta \frac{\partial L_{\rho}}{\partial\mathbf{Q}_n^t(k)},\label{equ:update_Q}\\ 
	&\mathbf{\Omega}_n^t \leftarrow \frac{K\left( \mathbf{Q}_n^{t\top}\mathbf{Q}_n^t\right)^{ \frac{1}{2} }}{tr\left(\left( \mathbf{Q}_n^{t\top}\mathbf{Q}_n^t\right)^{ \frac{1}{2} }\right)},\label{equ:update_O}\\
	&\mathbf{C}_{n} \leftarrow S_{1/\rho}\bigg( \mathbf{P}_{n}\mathbf{A} + \mathbf{S}_{n}\bigg), \label{equ:update_C} \\
	&\mathbf{S}_{n} \leftarrow \mathbf{S}_{n} + \mathbf{P}_{n}\mathbf{A} -\mathbf{C}_{n}, \label{equ:update_S}\\ 
	&\mathbf{D}_{n}(k) \leftarrow S_{1/\rho}\bigg( \mathbf{Q}_{n}(k)\mathbf{A} + \mathbf{U}_{n}(k)\bigg), \label{equ:update_D} \\
	&\mathbf{U}_{n}(k) \leftarrow \mathbf{U}_{n}(k) + \mathbf{Q}_{n}(k)\mathbf{A} -\mathbf{D}_{n}(k), \label{equ:update_U}\\
	&\mathbf{E}^{t} \leftarrow S_{1/\rho}\bigg( \mathbf{P}^{t}\mathbf{B} + \mathbf{V}^{t}\bigg), \label{equ:update_E} \\  
	&\mathbf{V}^{t} \leftarrow \mathbf{V}^{t} + \mathbf{P}^{t}\mathbf{B} - \mathbf{E}^{t},  \label{equ:update_V}\\
	&\mathbf{F}^{t}(k) \leftarrow S_{1/\rho}\bigg( \mathbf{Q}^{t}(k)\mathbf{B} + \mathbf{Z}^{t}(k)\bigg), \label{equ:update_F} \\  
	&\mathbf{Z}^{t}(k) \leftarrow \mathbf{Z}^{t}(k) + \mathbf{Q}^{t}(k)\mathbf{B} - \mathbf{F}^{t}(k),  \label{equ:update_Z} 
	\end{align} 
\end{small}
where $\eta$ is the learning rate of gradient descent. The derivative of $L_{\rho}$ with respect to $\mathbf{P}_n^t$ is:
\begin{small}
\begin{equation}\label{equ:derivative_P}
\begin{aligned}
\frac{\partial L_{\rho}}{\partial \mathbf{P}_n^t} & = 2\sum_{k=1}^{K}\left(\mathbf{X}_n^t \left(\mathbf{P}_n^t + \mathbf{Q}_n^t(k)\right) - Y_n^t(k)\right)\cdot\mathbf{X}_n^{t\top} \\
& + \rho (\mathbf{P}_{n}\mathbf{A} - \mathbf{C}_{n} + \mathbf{S}_{n})\cdot\mathbf{A}^{t\top} \\
& + \rho (\mathbf{P}^{t}\mathbf{B} -\mathbf{E}^{t} + \mathbf{V}^{t})\cdot\mathbf{B}_n^{\top},
\end{aligned}
\end{equation}
\end{small}
where $\mathbf{A}^t$ is the $t^{th}$ row of $\mathbf{A}$,  $\mathbf{B}_n$ is the $n^{th}$ row of $\mathbf{B}$. The derivative of $L_{\rho}$ with respect to $\mathbf{Q}_n^t(k)$ is:
\begin{small}
\begin{equation}\label{equ:derivative_Q}
\begin{aligned}
\frac{\partial L_{\rho}}{\partial \mathbf{Q}_n^t(k)} & = 2\left(\mathbf{X}_n^t \left(\mathbf{P}_n^t + \mathbf{Q}_n^t(k)\right) - Y_n^t(k)\right)\cdot\mathbf{X}_n^{t\top} \\
& + \alpha\left(\mathbf{Q}_{n}^{t}{\mathbf{\Omega}_n^t}^{-1}\right)(k)\\
& + \rho (\mathbf{Q}_{n}(k)\mathbf{A} - \mathbf{D}_{n}(k) + \mathbf{U}_{n}(k))\cdot\mathbf{A}^{t\top} \\
& + \rho (\mathbf{Q}^{t}(k)\mathbf{B} -\mathbf{F}^{t}(k) + \mathbf{Z}^{t}(k))\cdot\mathbf{B}_n^{\top},
\end{aligned}
\end{equation}
\end{small}
where $\big(\mathbf{Q}_{n}^{t}{\mathbf{\Omega}_n^t}^{-1}\big)(k)$ is the $k^{th}$ column of $\mathbf{Q}_{n}^{t}{\mathbf{\Omega}_n^t}^{-1}$. The \textit{soft thresholding operator} $S_{1/\rho}(x)$ is defined as follows:
\begin{small}
\begin{equation}\label{equ:soft_thresholding}
	S_{1/\rho}(x) = \begin{cases}
		x-1/\rho \,\,\,\,\,\,\,\,\,\,\,\,\,\,\,\,\,\,\, if\,\,\,\,\,\,x\,\,\,>\,\,\,1/\rho\\
		0 \,\,\,\,\,\,\,\,\,\,\,\,\,\,\,\,\,\,\,\,\,\,\,\,\,\,\,\,\,\,\,\,\,\,\,\,if\,\,\,\|x\| \leq\,\,\,1/\rho\\
		x+1/\rho \,\,\,\,\,\,\,\,\,\,\,\,\,\,\,\,\,\,\, if\,\,\,\,\,\,x\,\,\,<-1/\rho\\
	\end{cases}
\end{equation}
\end{small}
The details of ADMM optimization procedure are shown in Algorithm \ref{alg:admm}. We first initialize weight matrices $\mathbf{P}$ and $\mathbf{Q}$, auxiliary variable matrices $\mathbf{C}$, $\mathbf{D}$, $\mathbf{E}$ and $\mathbf{F}$, and scaled dual variables matrices $\mathbf{S}$, $\mathbf{U}$, $\mathbf{V}$ and $\mathbf{Z}$ randomly (line 1). Note that we initialize $\mathbf{\Omega} = \frac{1}{K}\mathbf{I}_K$ according to the assumption that all types of crime are unrelated initially. In each iteration of ADMM, we first leverage Gradient Descent technique with the gradient in Eq.~(\ref{equ:derivative_P}) and Eq.~(\ref{equ:derivative_Q}) to update the current $\mathbf{P}_n^t$ and $\mathbf{Q}_n^t(k)$ (line 4 and 6). Note that all $\mathbf{P}_{n'}^{t'}$ and $\mathbf{Q}_{n'}^{t'}(k')$ $(n'\not= n \ or \ t' \not= t \ or \ k' \not= k)$ are fixed. Then we update $\mathbf{\Omega}_n^t$ according to Eq.~(\ref{equ:update_O}) in line 8. Next we proceed to update $\mathbf{C}_n$, $\mathbf{S}_n$, $\mathbf{D}_n(k)$, $\mathbf{U}_n(k)$, $\mathbf{E}^t$, $\mathbf{V}^t$, $\mathbf{F}^t(k)$ and $\mathbf{Z}^t(k)$ using aforementioned update rules from line 10 to line 25. When ADMM optimization approaches convergence, Algorithm \ref{alg:admm} will output the well trained weight vectors $\mathbf{P}_n^t$ and $\mathbf{Q}_n^t(k)$, for $n \in [1,N],t \in [1,T],k \in [1,K]$ respectively.
\begin{figure*}
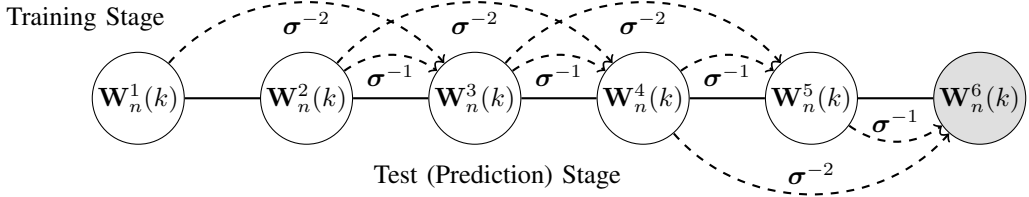

	\centering
	\tikz{ %
		\node[latent] (w1) {$\mathbf{W}_n^1(k)$} ; %
		\node (ui1) [above=0.15cm of w1,xshift=-0.7cm] {Training Stage};
		\node[latent, right=of w1] (w2) {$\mathbf{W}_n^2(k)$} ; %
		\node[latent, right=of w2] (w3) {$\mathbf{W}_n^3(k)$} ; %
		\node (ui3) [below=0.15cm of w3,xshift=0.3cm] {Test (Prediction) Stage};
		\node[latent, right=of w3] (w4) {$\mathbf{W}_n^4(k)$} ; %
		\node[latent, right=of w4] (w5) {$\mathbf{W}_n^5(k)$} ; %
		\node[obs, right=of w5] (w6) {$\mathbf{W}_n^6(k)$} ; %
		\edge[-,thick]{w1}{w2};
		\edge[-,thick]{w2}{w3};
		\edge[-,thick]{w3}{w4};
		\edge[-,thick]{w4}{w5};
		\edge[-,thick]{w5}{w6};
		\draw[->,dashed,thick] (w1) to [bend left=49] node[anchor=north]{$\boldsymbol \sigma^{-2}$} (w3);
		\draw[->,dashed,thick] (w2) to [bend left=36] node[anchor=north]{$\boldsymbol \sigma^{-1}$} (w3);		
		\draw[->,dashed,thick] (w2) to [bend left=49] node[anchor=north]{$\boldsymbol \sigma^{-2}$} (w4);
		\draw[->,dashed,thick] (w3) to [bend left=36] node[anchor=north]{$\boldsymbol \sigma^{-1}$} (w4);
		\draw[->,dashed,thick] (w3) to [bend left=49] node[anchor=north]{$\boldsymbol \sigma^{-2}$} (w5);
		\draw[->,dashed,thick] (w4) to [bend left=36] node[anchor=north]{$\boldsymbol \sigma^{-1}$} (w5);
		\draw[->,dashed,thick] (w4) to [bend right=49] node[anchor=south]{$\boldsymbol \sigma^{-2}$} (w6);
		\draw[->,dashed,thick] (w5) to [bend right=36] node[anchor=south]{$\boldsymbol \sigma^{-1}$} (w6);
	}
	\caption{An example of learning parameter $\sigma$ for crime prediction.\label{fig:test}}
%	\vspace{-4.9mm}
\end{figure*}

\begin{small}
\begin{algorithm}
	\caption{\label{alg:admm} The ADMM Optimization of CCC model.}
	\raggedright
	{\bf Input}: The feature vectors $\mathbf{X}$, the observed crime amounts $\mathbf{Y}$, the sparse matrices $\mathbf{A}$ and $\mathbf{B}$, parameter $\rho$\\
	{\bf Output}: The weight matrices $\mathbf{P}_n^t$ and $\mathbf{Q}_n^t(k)$, $\forall n \in [1,N]\,\forall t \in [1,T]\,\forall k \in [1,K]$\\
	\begin{algorithmic} [1]
		\STATE Initialize $\mathbf{P},\mathbf{Q},\mathbf{C},\mathbf{D},\mathbf{E},\mathbf{F},\mathbf{S},\mathbf{U},\mathbf{V},\mathbf{Z}$ randomly and initialize $\mathbf{\Omega} = \frac{1}{K}\mathbf{I}_K$,$\forall n \in [1,N]\,\forall t \in [1,T]\,\forall k \in [1,K]$
		\WHILE{Not Convergent}
		\FOR{$n \in [1,N],t \in [1,T]$}
		\STATE Calculate $\frac{\partial L_{\rho}}{\partial \mathbf{P}_n^t}$ according Eq. (\ref{equ:derivative_P}) and update $\mathbf{P}_n^t$ according to Eq. (\ref{equ:update_P})
		\FOR{$k \in [1,K]$}
		\STATE Calculate $\frac{\partial L_{\rho}}{\partial \mathbf{Q}_n^t(k)}$ according Eq. (\ref{equ:derivative_Q}) and update $\mathbf{Q}_n^t(k)$ according to Eq. (\ref{equ:update_Q})
		\ENDFOR
		\STATE Update $\mathbf{\Omega}_n^t$ according to Eq. (\ref{equ:update_O})
		\ENDFOR
		\FOR{$n \in [1,N]$}
		\STATE Update $\mathbf{C}_n$ according to Eq. (\ref{equ:update_C})
		\STATE Update $\mathbf{S}_n$ according to Eq. (\ref{equ:update_S})
		\FOR{$k \in [1,K]$}
		\STATE Update $\mathbf{D}_n(k)$ according to Eq. (\ref{equ:update_D})
		\STATE Update $\mathbf{U}_n(k)$ according to Eq. (\ref{equ:update_U})
		\ENDFOR
		\ENDFOR
		\FOR{$t \in [1,T]$}
		\STATE Update $\mathbf{E}^t$ according to Eq. (\ref{equ:update_E})
		\STATE Update $\mathbf{V}^t$ according to Eq. (\ref{equ:update_V})
		\FOR{$k \in [1,K]$}
		\STATE Update $\mathbf{F}^t(k)$ according to Eq. (\ref{equ:update_F})
		\STATE Update $\mathbf{Z}^t(k)$ according to Eq. (\ref{equ:update_Z})
		\ENDFOR
		\ENDFOR
		\ENDWHILE
	\end{algorithmic}
\end{algorithm}
\end{small}

Next we discuss the computational cost of Algorithm~\ref{alg:admm}. In each iteration of ADMM, calculating $\frac{\partial L_{\rho}}{\partial \mathbf{Q}_n^t}$ according to Eq.~(\ref{equ:derivative_Q}) is the most time consuming step. First we consider the time complexity of the first term in Eq.~(\ref{equ:derivative_Q}), in which $\mathbf{X}_n^t \mathbf{P}_n^t$ and $\mathbf{X}_n^t \mathbf{Q}_n^t(k)$ can be computed in $O(M^2)$, then  subtracting $Y_n^t(k)$ and multiplying $\mathbf{X}_n^{t\top}$ can be computed in $O(M)$, so the time complexity of the first term is $O(M^2 + M)$. The second term can be computed in $O(M*K^2)$. For the third term, since the matrix representation of $\mathbf{A}$ is very sparse, i.e., each row or column of $\mathbf{A}$ has at most two non-zero elements, thus the time complexity of it is $O(M*T)$. Then the multiplying $\mathbf{A}^{t\top}$ can be computed in $O(M*T)$, so time complexity of the third term is $O(M*T)$. Similarly, the last term can be computed in $O(M*N)$. Therefore, considering that there are $N$ regions, $T$ time slots and $K$ types of crime, the computational cost of each ADMM iteration is $O(NTK(M^2 + M*K^2 + M*T + M*N))$. %Note that it is straightforward to be parallelized for large-scale datasets since our optimization method is based on ADMM. 

\subsection{Crime Prediction Task}
\label{sec:prediction}
When ADMM is convergent, Algorithm \ref{alg:admm} can output the well trained weight vectors $\mathbf{P}_n^t$ and $\mathbf{Q}_n^t(k)$, for all $n \in [1,N],t \in [1,T],k \in [1,K]$. In this subsection, we introduce how to perform crime prediction for a future time slot (i.e. $(T+1)^{th}$ time slot) based on all $\mathbf{P}_n^t$ and $\mathbf{Q}_n^t(k)$.

As mentioned in Section~\ref{sec:problem}, we actually construct feature vector $\mathbf{X}_n^t$ using data in $(t-1)^{th}$ time slot rather than $t^{th}$ time slot of $n^{th}$ region. Thus for the $(T+1)^{th}$ time slot, we can construct $\mathbf{X}_n^{T+1}$ based on data in $T^{th}$ time slot. Therefore, in order to predict crime amount $Y_n^{T+1}(k) = \mathbf{X}_n^{T+1} \left(\mathbf{P}_n^{T+1} + \mathbf{Q}_n^{T+1}(k)\right)$ for $k^{th}$ type of crime in $n^{th}$ region of $T+1^{th}$ time slot, we need the mapping vectors $\mathbf{P}_n^{T+1}$ and $\mathbf{Q}_n^{T+1}(k)$. To sum up, the problem becomes to estimate $\mathbf{P}_n^{T+1}$ and $\mathbf{Q}_n^{T+1}(k)$ based on $ \{\mathbf{P}_n^{t} \}_{t=1}^T$ and $ \{\mathbf{Q}_n^{t}(k) \}_{t=1}^T$.

The mapping vectors $\mathbf{P}_n^{t}$ and $\mathbf{Q}_n^{t}(k)$ should be related to these of previous time slots according to intra-region temporal correlation. Therefore, we assume that $\mathbf{W}_n^{t}(k)= \mathbf{P}_n^{t} + \mathbf{Q}_n^{t}(k)$ is the weighted sum of its previous $\mathcal{G}$ time slots as:
%\begin{equation}
%\begin{aligned}
%\hat{\mathbf{P}_n^{t}} & = \frac{\sum_{\Delta t=1}^{\mathcal{G}} f(\Delta t)\mathbf{P}_n^{t-\Delta t}}{\sum_{\Delta t=1}^{\mathcal{G}} f(\Delta t)}\\
%& = \frac{f(1)\mathbf{P}_n^{t-1} + f(2)\mathbf{P}_n^{t-2} + \ldots + f(\mathcal{G})\mathbf{P}_n^{t-\mathcal{G}}}{\sum_{\Delta t=1}^{\mathcal{G}} f(\Delta t)}
%\end{aligned}
%\end{equation}
\begin{small}
\begin{equation}
\begin{aligned}
\mathbf{W}_n^{t}(k) = &\frac{\sum_{\Delta t=1}^{\mathcal{G}} f(\Delta t)\left(\mathbf{P}_n^{t-\Delta t} + \mathbf{Q}_n^{t-\Delta t}(k)\right)}{\sum_{\Delta t=1}^{\mathcal{G}} f(\Delta t)}\\
= & \frac{f(1)\left(\mathbf{P}_n^{t-1} + \mathbf{Q}_n^{t-1}(k)\right) + \ldots + f(\mathcal{G})\left(\mathbf{P}_n^{t-\mathcal{G}} + \mathbf{Q}_n^{t-\mathcal{G}}(k)\right)}{f(1) + \ldots + f(\mathcal{G})},
\end{aligned}
\end{equation}
\end{small}
where $f(\Delta t)$ should be a non-increase function of $\Delta t$, i.e., $f(\Delta t)$ should be larger when $\Delta t$ is smaller, since $\mathbf{W}_n^{t}$ should be closer related to its just previous few time slots. In this work, we use a power law exponential function of $f(\Delta t) = \sigma^{-\Delta t}$, where $\sigma \in [1,+ \infty)$ is introduced to control the contributions from $\{\mathbf{W}_n^{t-1},\mathbf{W}_n^{t-2},$ $\ldots, \mathbf{W}_n^{t-\mathcal{G}} \}$. Note that when $\sigma = 1$, $\{\mathbf{W}_n^{t-1},\mathbf{W}_n^{t-2}, \ldots, \mathbf{W}_n^{t-\mathcal{G}} \}$ contributes equally to $\mathbf{W}_n^{t}$. We propose to automatically estimate optimal $\sigma$ from the training data via solving the following optimization problem:
\begin{small}
\begin{equation}\label{eq:parameter-train}
\min_{\sigma}  \sum_{t=\mathcal{G}+1}^T \left( \mathbf{X}_n^{t} \frac{\sum_{\Delta t=1}^{\mathcal{G}} \sigma^{-\Delta t}\left(\mathbf{P}_n^{t-\Delta t}+ \mathbf{Q}_n^{t-\Delta t}(k)\right)}{\sum_{\Delta t=1}^{\mathcal{G}} \sigma^{-\Delta t}} -Y_n^{t}(k) \right)^2.
\end{equation}
\end{small}

Figure \ref{fig:test} illustrates how we learn $\sigma$ for $k^{th}$ type of crime in $n^{th}$ region, where we use $\mathbf{W}_n^t(k) = \mathbf{P}_n^t + \mathbf{Q}_n^t(k)$. In this example, we aim to predict crime amount in $6^{th}$ time slot based on the training data of previous $T=5$ time slots and the well trained weight vectors $\{\mathbf{W}_n^1(k), \mathbf{W}_n^2(k), \mathbf{W}_n^3(k), \mathbf{W}_n^4(k), \mathbf{W}_n^5(k)\}$ from Algorithm \ref{alg:admm}. We use previous $\mathcal{G}=2$ time slots to predict ${\bf W}_n^6(k)$ as $\mathbf{W}_n^{6}(k) = \frac{\sigma^{-1}\mathbf{W}_n^{5}(k) + \sigma^{-2} \mathbf{W}_n^{4}(k)}{\sigma^{-1} + \sigma^{-2}}$. By solving Eq.~(\ref{eq:parameter-train}), we can estimate $\sigma$ based on $T-\mathcal{G} = 3$ samples, i.e., $\mathbf{W}_n^{5}(k) = \frac{\sigma^{-1}\mathbf{W}_n^{4}(k) + \sigma^{-2} \mathbf{W}_n^{3}(k)}{\sigma^{-1} + \sigma^{-2}}$, $\mathbf{W}_n^{4}(k) = \frac{\sigma^{-1}\mathbf{W}_n^{3}(k) + \sigma^{-2} \mathbf{W}_n^{2}(k)}{\sigma^{-1} + \sigma^{-2}}$ and $\mathbf{W}_n^{3}(k) = \frac{\sigma^{-1}\mathbf{W}_n^{2}(k) + \sigma^{-2} \mathbf{W}_n^{1}(k)}{\sigma^{-1} + \sigma^{-2}}$. Then the number of $k^{th}$ type of crime in $6^{th}$ time slot can be predicted as: $\hat{Y_n^6(k)} = {\bf X}_n^6\frac{\sigma^{-1}\mathbf{W}_n^{5}(k) + \sigma^{-2} \mathbf{W}_n^{4}(k)}{\sigma^{-1} + \sigma^{-2}}$. Finally, it worth to note that different types of crime in different regions may have different temporal patterns. Thus we learn the parameters $\sigma_n(k)$ to estimate $\mathbf{W}_n^{T+1}(k)$ for each type of crime in each region, respectively. 
\section{Experiments}
\label{sec:experiments}
In this section, we conduct extensive experiments to evaluate the effectiveness of the proposed framework. We first introduce the urban data and experimental settings. Then we seek to answer two questions: (1) how the proposed framework performs compared to the state-of-the-art baselines; and (2) how the cross-type correlations and spatio-temporal correlations benefit crime prediction. Finally, we investigate how the important parameters affect the performance of crime prediction. 

\subsection{Data}
\label{sec:data_and_features}
The data of $K=7$ types of crime is collected from 07/01/2012 to 06/30/2013 ($T=365$ days) in New York City. We segment NYC into $N=100$ disjointed $2km \times 2km$ grids (regions). For the feature matrices, we collect multiple data resources that are related to crime. 
%The filtered data and implementation code is available online\footnote{\url{https://www.dropbox.com/s/lkfqyw4jqdswjts/Crime.zip?dl=0}}. 
Then we detail these resources. 
\begin{itemize}[leftmargin=*]
	\item \textbf{Crime Complaint Data}\footnote{\url{https://data.cityofnewyork.us/Public-Safety/NYPD-Complaint-Data-Historic/qgea-i56i/data}}: Regions with many crime complaints tend to occur more crimes in the near future. Thus we collect crime complaint data with complaint frequencies of the aforementioned 7 types of crimes. 
	\item \textbf{Stop-and-Frisk Data}\footnote{\url{https://www1.nyc.gov/site/nypd/stats/reports-analysis/stopfrisk.page}}: Stop-and-Frisk is a crime prevention program of NYC police department that temporarily detains, questions and searches citizens for weapons on the street. We collect stop-and-frisk dataset since this program is claimed that contributes to the decline of urban crimes~\cite{weisburd2016stop}. 
	\item \textbf{Meteorological Data}\footnote{We crawl the meteorological data via \url{http://api.wunderground.com/}}: Crime is strongly associated with meteorology~\cite{cohn1990weather}. Therefore, we collect meteorological dataset containing $30$ features like weather, temperature, pressure, wind strength, precipitation, etc. 
	\item \textbf{Point of interests (POIs) Data}\footnote{\url{https://sites.google.com/site/yangdingqi/home/foursquare-dataset}}: POIs represent the function of regions, which could benefit crime prediction. We crawled 10 types of POIs, i.e., food, shop, residence, nightlife, entertainment, travel, outdoors, professional, education and event from FourSquare.
	\item \textbf{Human Mobility Data}:  Human mobility provides useful information like residential stability and population density, which is related to urban crime. We extract human check-ins from FourSquare\footnote{\url{https://sites.google.com/site/yangdingqi/home/foursquare-dataset}}, and taxi pick-up\&drop-off points from the taxi GPS data\footnote{\url{https://www1.nyc.gov/site/tlc/about/tlc-trip-record-data.page}}. 
	\item \textbf{311 Public-Service Complaint Data}\footnote{\url{https://data.cityofnewyork.us/Social-Services/311-Service-Requests-from-2010-to-Present/erm2-nwe9/data}}: 311 is the service number of NYC government, which allows citizens to complain about things like electric, water, traffic, etc. 311 reveals citizens' dissatisfaction with government service, which is related with urban crime.	 
\end{itemize}

%In this work, we focus on extracting features from the aforementioned sources. It is possible to also use other sources such as criminal networks, social media and urban configuration. We will leave it as one future work. 

\subsection{Experimental Settings}
\label{sec:experimental_settings}
For each type of crime in each region, we leverage previous $\mathcal{T}=7$ time slots' data to train the parameters since crime amounts are typically associated to recent previous time slots, and predict the crime amount of $\tau$ time slots later (we vary $\tau=\{1,7\}$). Thus, in each region, each type of crime has $T_S = T-\mathcal{T}-\tau+1$ test samples in total, where $T=365$ is the total number of time slots. 

The performance of crime prediction is evaluated in terms of the average root-mean-square-error (RMSE) of all $K$ types of crime in $N$ regions:
\begin{small}
\begin{equation}\label{equ:armse}
	RMSE = \frac{1}{NK}\sum_{n=1}^N\sum_{k=1}^K \sqrt{\frac{1}{T_S}\sum_{t_s=1}^{T_S}{\left( \hat {\mathbf{Y}}_n^{t_s}(k) - \mathbf{Y}_n^{t_s}(k)\right)^2}},
\end{equation}
\end{small}
where $\hat {\mathbf{Y}}_n^{t_s}(k)$ is the predicted crime amount and $\mathbf{Y}_n^{t_s}(k)$ is the observed number. We select parameters of the proposed framework such as $\alpha$, $\beta$, $\gamma$, $\rho$ and $\sigma$ by cross-validation. More details about parameter selection will be discussed in following subsections. 

\subsection{Performance Comparison for Crime Prediction}
\label{sec:ev_overall}
To seek answer of the first question, we compare the proposed framework with the state-of-the-art baseline methods. For a fair comparison, we conduct parameter-tuning for each baseline. Next, we detail the baselines as follows:
\begin{itemize}[leftmargin=*]
	\item \textbf{ARIMA}: Auto-Regression-Integrated-Moving-Average is used to for short-term crime prediction which considers the recent $\mathcal{T}$ days for a moving average in~\cite{chen2008forecasting}.  
	\item \textbf{VAR}: Vector Auto-Regression is a multi variate forecasting technique accounting for cross correlation and temporal correlation, which is leveraged to forecast crime in~\cite{corman1987crime}.	
	\item \textbf{RNN}: Recurrent Neural Network is to predict incidents such as murder and robbery in~\cite{cortez2018architecture}, where connections between units form a directed graph, which allows it to exhibit dynamic temporal behavior for a sequence.
	\item \textbf{DeepST}: DeepST is a DNN-based prediction model for spatio-temporal data~\cite{zhang2016dnn}. CNN is used to extract spatio-temporal properties from historical crime density maps, and meteorological data is used as the global information.
	\item \textbf{ST-ResNet}: ST-ResNet~\cite{zhang2017deep} is
	a deep learning based approach upon DeepST, where residual units are introduced to enhance training effectiveness of a very deep network.
	\item \textbf{stMTL}: Spatio-Temporal Multi-Task Learning enhances static spatial smoothness regression framework by learning the temporal dynamics of features through a non-parametric term~\cite{zheng2013time}.
	\item \textbf{TCP}: This baseline captures spatio-temporal correlations including intra-region temporal and the inter-region spatial correlations for crime prediction~\cite{zhao2017modeling}. 
\end{itemize}
%	\item \textbf{MLR}: Multiple Linear Regression is adopted for violent crime prediction in~\cite{gomez2004violent}, two or more independent variables are used to predict the value of a dependent variable. 
%	\item \textbf{LASSO}: Lasso Regression tries to minimize the objective function $\frac{1}{2}\|\mathbf{Y}^t - \mathbf{X}^t\mathbf{W}^t\|_2^2 + \gamma\|\mathbf{W}^t\|_1$ and encodes the sparsity over all weights in $\mathbf{W}^t$ for crime prediction in~\cite{kadar2016exploring}. 

\begin{figure}[t]
	\centering
	\includegraphics[width=81mm]{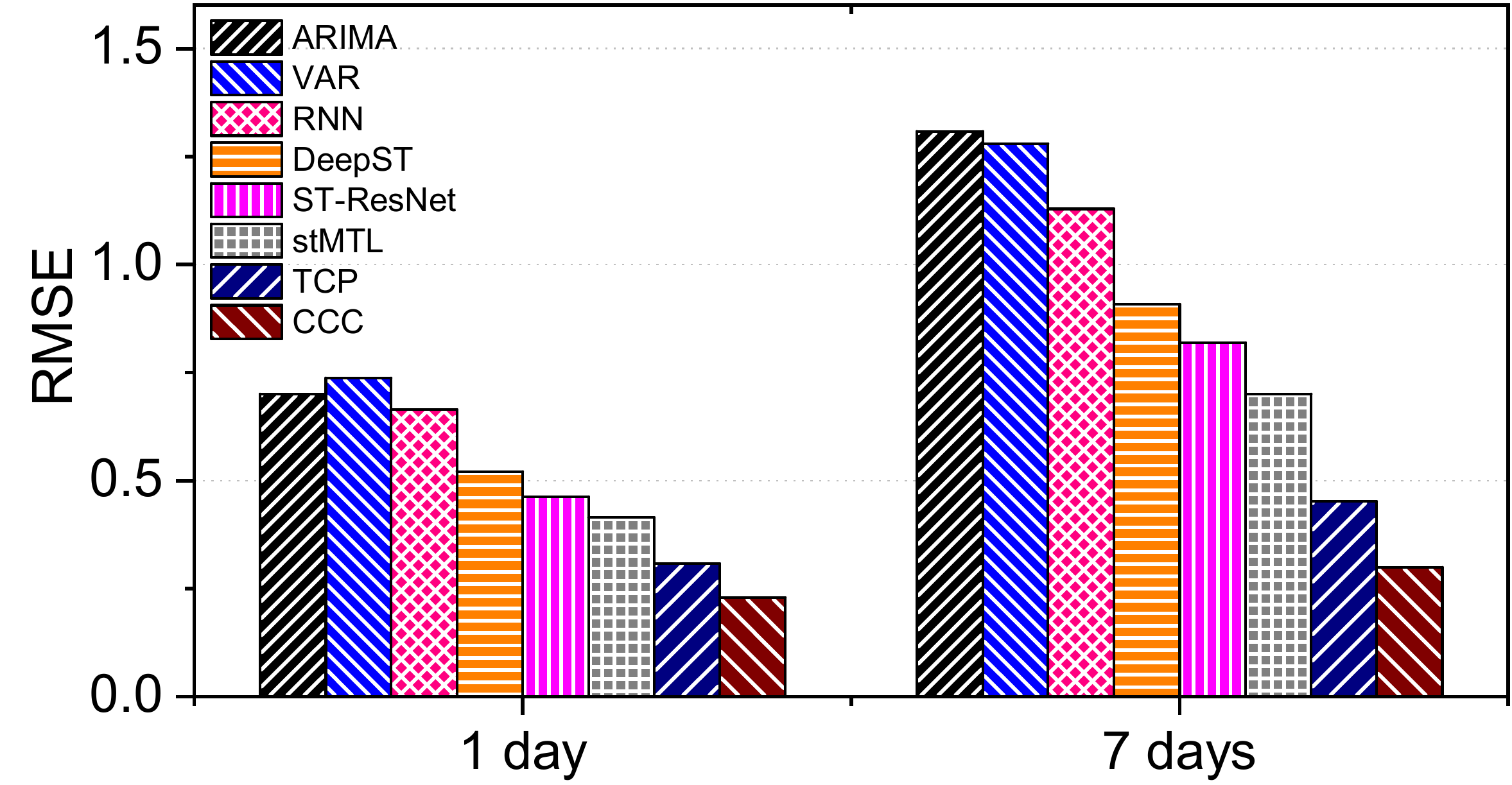}
	\caption{Overall performance comparison in terms of RMSE.}
	\label{fig:overall}
%	\vspace{-4.9mm}
\end{figure}

The results are shown in Figure~\ref{fig:overall}. Note that we leverage cross-validation to tune the parameters in baselines and our framework. We have following observations:
\begin{itemize}[leftmargin=*]
	\item DeepST and ST-ResNet outperform the previous three methods, which demonstrates that the crimes among different regions are indeed spatially correlated. The first three baselines only consider the temporal dependencies, while overlook the spatial correlations. %RNN outperforms ARIMA and VAR, which demonstrates the advancement of deep neural networks over traditional models.
	\item stMTL and TCP achieve the better performance than previous five methods, since stMTL and TCP incorporate multiple sources that are related to crime, while previous five methods predict the crime amount solely based on the historical crime records (DeepST and ST-ResNet also incorporate meteorological data). TCP performs better than stMTL, because stMTL only captures features cross regions in the same time slot that share the same weights; while TCP captures both spatio-temporal correlations. 
	\item CCC performs better than TCP, because CCC jointly captures cross-type correlations among multiple types of crime and spatio-temporal correlations for each type of crime, while TCP overlooks the cross-type correlations.
	\item All methods perform relatively better in short-term (1 day) crime prediction, which indicates that prediction of distant future is harder than that of near future. However, the proposed framework performs more robustly in long-term (7 days) prediction than baseline techniques. 
\end{itemize}

According to above observations, we can answer the first question -- the proposed CCC framework can outperform the state-of-the-art baselines for crime prediction by introducing cross-type and spatio-temporal correlations among multiple types of crime.

\subsection{Contributions of Important Components}
\label{sec:ev_components}
In this subsection, we study the contribution of each important component of the proposed framework. We systematically eliminate each component and define following variants of CCC: 
\begin{itemize}[leftmargin=*]
	\item \textbf{CCC}$-c$: In this variant, we evaluate the contribution of cross-type correlation, so we eliminate the impact from cross-type correlation by setting $\alpha = 0$.
	\item \textbf{CCC}$-t$: This variant is to evaluate the performance of intra-region temporal correlations, so we set parameters of temporal correlation as 0, i.e., $\beta = 0$.  
	\item \textbf{CCC}$-s$: In this variant, we evaluate the contribution of inter-region spatial correlation, so we eliminate the impact from it by setting all $d(i,j)^{-\gamma}$ as $0$.
	\item \textbf{CCC}$-p$: This variant is to evaluate the performance of weight $\mathbf{P}$ that captures the common features for all types of crime, so we remove all $\mathbf{P}_n^t$ for $n \in [1,N], k \in [1,K]$.
\end{itemize}

\begin{figure}[t]
	\centering
	\includegraphics[width=81mm]{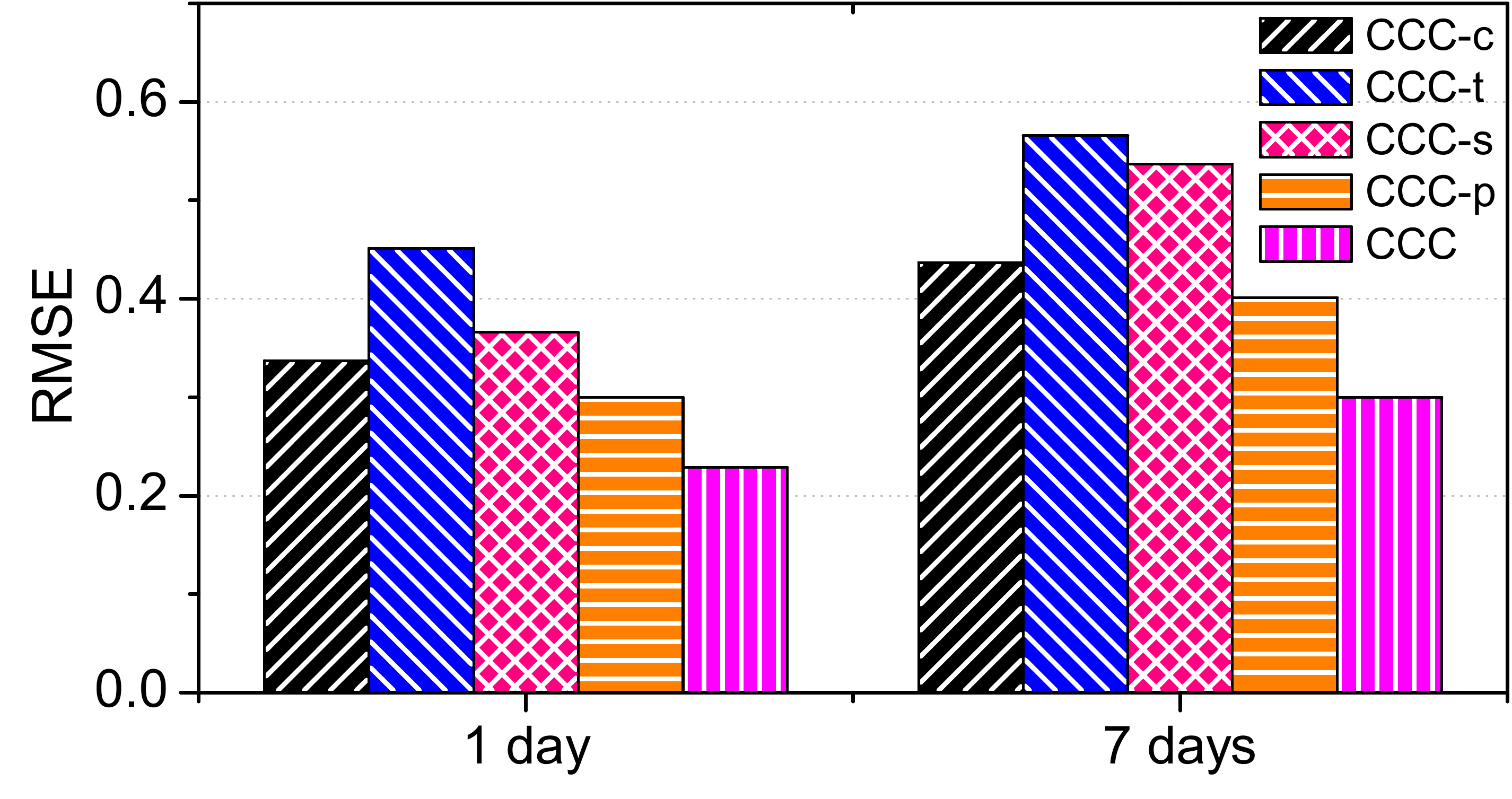}
	\caption{Impact of important correlations and components.}
	\label{fig:Components}
%	\vspace{-4.9mm}
\end{figure}

The results are shown in Figure~\ref{fig:Components}. From this figure, we can observe:
\begin{itemize}[leftmargin=*]
	\item CCC achieves better performance than CCC$-c$ in both 1-day and 7-day prediction, which verifies that different types of crime are intrinsically correlated and introducing cross-type correlations can boost the performance of crime prediction.
	\item CCC$-s$ outperforms CCC$-t$ in both 1-day and 7-day prediction, which shows that intra-region temporal correlation contributes more in crime prediction. Their performance becomes close in 7-day prediction. This indicates that temporal correlation becomes weak in long-term prediction.
	\item CCC performs better than CCC$-p$. This result supports that introducing weight vectors $\mathbf{P}$ to capture the common features for all types of crime is helpful for crime prediction.
\end{itemize}

To sum up, we can answer the second question - CCC outperforms all its variants, which supports that all components are useful in crime prediction and they contain complementary information. 

\subsection{Parametric Sensitivity Analysis}
\label{sec:ev_parametric}
In this section, we evaluate three key parameters of the proposed framework, i.e., (1) $\alpha$ that controls cross-type correlation, (2) $\beta$ that controls temporal correlation, and (3) $\gamma$ that controls spatial correlation. To investigate the sensitivity of the proposed framework CCC with respect to these parameters, we study how CCC performs with changing the value of one parameter, while keeping other parameters fixed.

\begin{figure}[t]
	\centering
	\includegraphics[width=81mm]{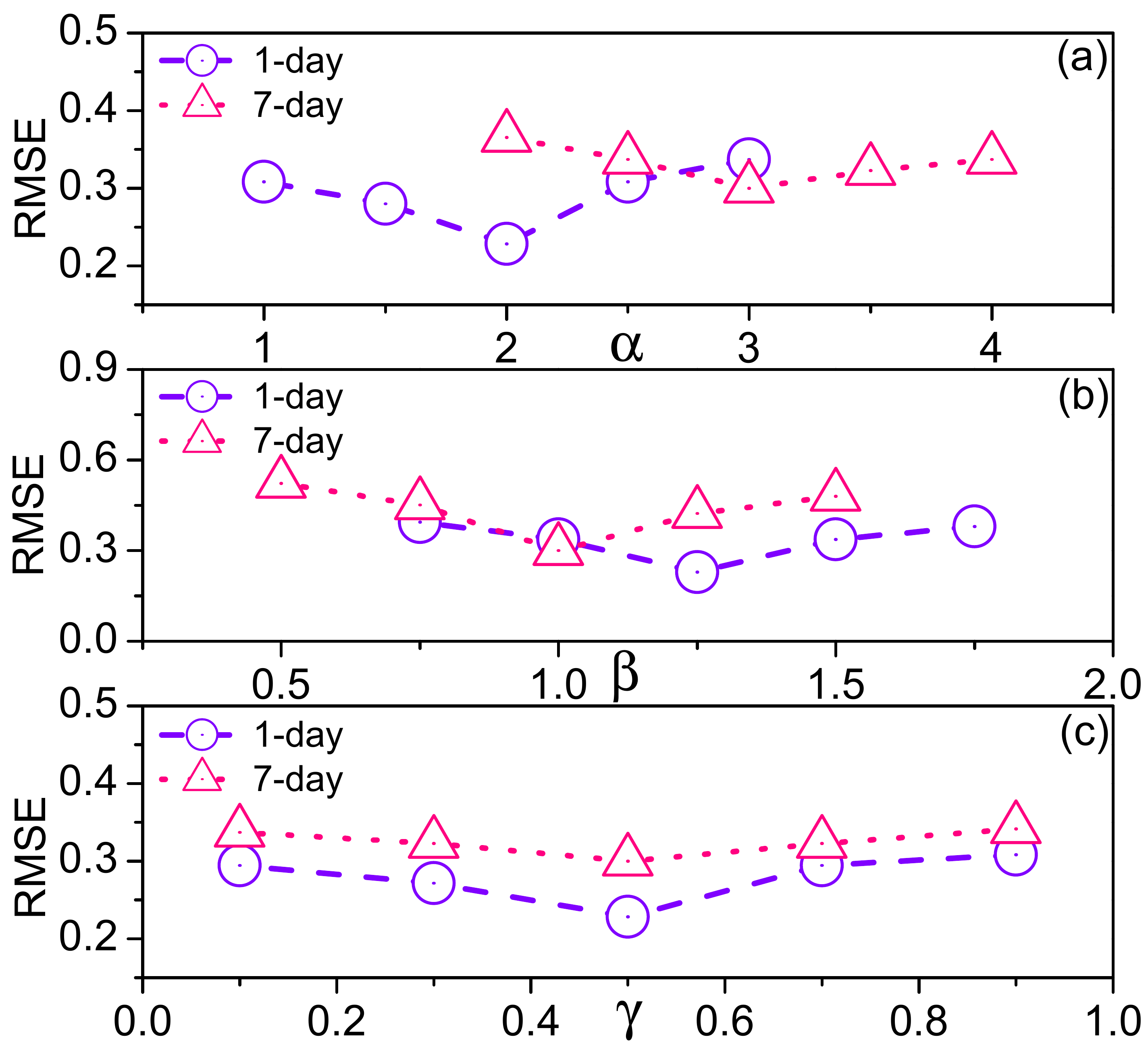}
	\caption{Parameter sensitiveness. (a) $\alpha$ for cross-type correlation, (b) $\beta$ for temporal correlation, (c) $\gamma$ for spatial correlation.}
	\label{fig:parameters_sensitivity}
%	\vspace{-4.9mm}
\end{figure}

Figure~\ref{fig:parameters_sensitivity} (a) illustrates the parameter sensitivity of $\alpha$ for crime prediction. The proposed framework achieves the best performance when $\alpha=2$ for 1-day prediction, while $\alpha=3$ for 7-day prediction. This result indicates that cross-type correlation plays a more important role in long-term prediction.

For temporal correlation, Figure~\ref{fig:parameters_sensitivity} (b) shows how the performance changes with $\beta$. The performance achieves the peak when $\beta = 1.25$ for 1-day prediction and $\beta = 1$ for 7-day prediction, which suggests that weight vectors $\mathbf{P}_n^{t}$ and $\mathbf{Q}_n^{t}(k)$ are closely related to these of just the last few time slots; while in distant future prediction, the temporal correlation becomes weak.

Figure \ref{fig:parameters_sensitivity} (c) shows the parameter sensitivity of $\gamma$. When $\gamma \to 0, d(i,j)^{-\gamma} \to 1$, i.e., all regions are equally related to each other, or $\gamma \to +\infty, d(i,j)^{-\gamma} \to 0$, i.e., all regions are independent of each other. CCC approaches the best performance when $\gamma = 0.5$ for both 1-day and 7-day prediction, which demonstrates the importance of spatial correlation in crime prediction. 

\section{Related Work}
\label{sec:related_work}
In this section, we briefly introduce the current crime prediction techniques related to our study.  Typically, current techniques can be classified into three groups. 

The first group of techniques is based on statistical methods.  For example, researchers show that there is correlation between the characteristics of a population and the rate of violent crimes \cite{gruenewald2006ecological}.  The author in \cite{featherstone2013identifying} is able to discover a correlation between reported crime census statistics from the South African Police Service and crime events discussed in tweets. While authors in \cite{green2006symbolic} conclude that there is a positive effect of symbolic racism on both preventive and punitive penalties. Some researchers studied the trend of using web-based crime mapping from the 100 highest GDP cities of the world \cite{leong2013content}. %The authors conclude that the main factors that drive numerous crime mapping are e-governance and community policing. Another work proposes a novel Bayesian based prediction model to predict the accurate location of the next crime scene in a serial crime \cite{liao2010novel}.

The second group of techniques is data mining methods. For instance, in \cite{zhao2017exploring}, spatio-temporal patterns in urban data are exploited in one borough in New York City, and then the authors leverage transfer learning techniques to reinforce the crime prediction of other boroughs. Another researcher built a crime policing self-organizing map to extract information such as crime type and location from reports to provide for a more effective crime analysis and employs an unsupervised Sequential Minimization Optimization for clustering \cite{alruily2012using}. A four-order tensor for crime forecasting is presented in~\cite{mu2011empirical}. The tensor encodes the longitude, latitude, time, and other related crimes. In~\cite{yu2014crime}, a new feature selection and construction method are proposed for crime prediction by using temporal and spatial patterns. %the author in \cite{gerber2014predicting} uses Latent Dirichlet Allocation for learning topics and related terms from tweets and eschews deep semantic analysis in favor of shallower analysis via topic modeling. I

Finally, the third group of methods predicts crime by seismic analysis techniques. For instance, temporal patterns of dynamics of violence are analyzed using a point process model for crime prediction~\cite{lewis2012self}. In~\cite{mohler2011self}, self-exciting point process models are implemented for predicting crimes, which leverages a nonparametric evaluation strategy to gain an understanding of temporal tendencies for burglary. 
\section{Conclusion}
\label{sec:conclusion}
In this paper, we propose a novel framework CCC, which jointly captures cross-type and spatio-temporal correlations for crime prediction. CCC leverages heterogeneous big urban data, e.g., crime complaint data, stop-and-frisk data, meteorological data, point of interests (POIs) data, human mobility data and 311 public-service complaint data. We evaluate our framework with extensive experiments based on real-world urban data from New York City. The results show that (1) different types of crime are intrinsically correlated with each other, (2) the proposed framework can accurately predict crime amounts in the near future and (3) cross-type and spatio-temporal correlations can boost crime prediction. 

There are several interesting research directions. First, in addition to the cross-type and spatio-temporal correlations we studied in this work, we would like to investigate more crime patterns (e.g. periodicity and tendency) and model them mathematically for accurate crime prediction. Second, we would like to introduce and develop more advanced techniques for crime analysis. Third, besides crime prediction task, we would like to design more sophisticated models to tackle more practical policing tasks in the real world. 
\bibliographystyle{IEEEtran}
\bibliography{10Reference}

%\appendix

\end{document}